\documentstyle[12pt]{article}
\begin{document}
\thispagestyle{empty}

\newcommand{\p}[1]{(\ref{#1})}
\newcommand{\be}{\begin{equation}}
\newcommand{\ee}{\end{equation}}
\newcommand{\sect}[1]{\setcounter{equation}{0}\section{#1}}

\newcommand{\vs}[1]{\rule[- #1 mm]{0mm}{#1 mm}}
\newcommand{\hs}[1]{\hspace{#1mm}}
\newcommand{\mb}[1]{\hs{5}\mbox{#1}\hs{5}}
\newcommand{\Db}{{\overline D}}
\newcommand{\bea}{\begin{eqnarray}}
\newcommand{\eea}{\end{eqnarray}}
\newcommand{\wt}[1]{\widetilde{#1}}
\newcommand{\und}[1]{\underline{#1}}
\newcommand{\ov}[1]{\overline{#1}}
\newcommand{\sm}[2]{\frac{\mbox{\footnotesize #1}\vs{-2}}
		   {\vs{-2}\mbox{\footnotesize #2}}}
\newcommand{\prt}{\partial}
\newcommand{\eps}{\epsilon}

\newcommand{\R}{\mbox{\rule{0.2mm}{2.8mm}\hspace{-1.5mm} R}}
\newcommand{\Z}{Z\hspace{-2mm}Z}

\newcommand{\cd}{{\cal D}}
\newcommand{\cg}{{\cal G}}
\newcommand{\ck}{{\cal K}}
\newcommand{\cw}{{\cal W}}

\newcommand{\vj}{\vec{J}}
\newcommand{\vl}{\vec{\lambda}}
\newcommand{\vz}{\vec{\sigma}}
\newcommand{\vt}{\vec{\tau}}
\newcommand{\vw}{\vec{W}}
\newcommand{\poiss}{\stackrel{\otimes}{,}}

\def\l#1#2{\raisebox{.2ex}{$\displaystyle
  \mathop{#1}^{{\scriptstyle #2}\rightarrow}$}}
\def\r#1#2{\raisebox{.2ex}{$\displaystyle
 \mathop{#1}^{\leftarrow {\scriptstyle #2}}$}}


\newcommand{\NP}[1]{Nucl.\ Phys.\ {\bf #1}}
\newcommand{\PL}[1]{Phys.\ Lett.\ {\bf #1}}
\newcommand{\NC}[1]{Nuovo Cimento {\bf #1}}
\newcommand{\CMP}[1]{Comm.\ Math.\ Phys.\ {\bf #1}}
\newcommand{\PR}[1]{Phys.\ Rev.\ {\bf #1}}
\newcommand{\PRL}[1]{Phys.\ Rev.\ Lett.\ {\bf #1}}
\newcommand{\MPL}[1]{Mod.\ Phys.\ Lett.\ {\bf #1}}
\newcommand{\BLMS}[1]{Bull.\ London Math.\ Soc.\ {\bf #1}}
\newcommand{\IJMP}[1]{Int.\ Jour.\ of\ Mod.\ Phys.\ {\bf #1}}
\newcommand{\JMP}[1]{Jour.\ of\ Math.\ Phys.\ {\bf #1}}
\newcommand{\LMP}[1]{Lett.\ in\ Math.\ Phys.\ {\bf #1}}

\renewcommand{\thefootnote}{\fnsymbol{footnote}}
\newpage
\setcounter{page}{0}
\pagestyle{empty}
\begin{flushright}
{January 1997}\\
{JINR E2-97-37}\\
{solv-int/9701020}
\end{flushright}
\vs{8}
\begin{center}
{\LARGE {\bf The discrete symmetries of the}}\\[0.6cm]
{\LARGE {\bf $N=2$ supersymmetric GNLS hierarchies}}\\[1cm]

\vs{8}

{\large A. Sorin$^{1}$}
{}~\\
\quad \\
{\em {Bogoliubov Laboratory of Theoretical Physics, JINR,}}\\
{\em 141980 Dubna, Moscow Region, Russia}~\quad\\

\end{center}
\vs{8}

\centerline{ {\bf Abstract}}
\vs{4}

The discrete symmetry transformations of the $N=2$ supersymmetric
$(n,m)$-GNLS hierarchy are constructed. Their bosonic limit is analyzed
and new discrete symmetries of the modified GNLS hierarchy are
derived. The explicit relations connecting the integrable hierarchy,
produced by the junction of the Lax operators for the $N=2$ supersymmetric
$a=4$ KdV and $(n-1,m)$-GNLS hierarchies, to the $N=2$ supersymmetric
$(n,m)$-GNLS hierarchy are established.

\vfill
1){\em e-mail: sorin@thsun1.jinr.ru}

\newpage
\pagestyle{plain}
\renewcommand{\thefootnote}{\arabic{footnote}}
\setcounter{footnote}{0}

\noindent{\bf 1. Introduction.}
The goal of the present Letter is to construct the mappings that
act like the discrete symmetry transformations of the $N=2$ supersymmetric
$(n,m)$ Generalized Nonlinear Schr\"{o}dinger ($(n,m)$-GNLS) hierarchy
\cite{bks}. Recently, a variety of $N=2$ supersymmetric integrable
hierarchies, derived by the junction of the Lax operators for the $N=2$
supersymmetric $(n-1,m)$-GNLS and $a=4$ KdV \cite{lm,kst} hierarchies, was
proposed in \cite{ik}. We also explain its origin. We demonstrate that
this variety is gauge related to the variety of $N=2$ supersymmetric
$(n,m)$-GNLS hierarchies.

Let us start with a short summary of the main facts concerning
the $N=2$ supersymmetric $(n,m)$-GNLS hierarchy \cite{bks} and introduce
some new relations which will be useful in what follows.

The Lax operator of the $N=2$ supersymmetric $(n,m)$-GNLS
hierarchy has the following form\footnote{Hereafter, summation over
repeated indices is understood and the square brackets mean that entering
operators act only on superfields inside the brackets, e.g., the fermionic
derivative $D$ in the Lax operator \p{suplax} acts only on the term
${\overline F}_a$ inside the brackets.}:
\begin{eqnarray}
L= \partial - \frac{1}{2}(F_a {\overline F}_a +
F_a {\overline D}\partial^{-1} \left[ D {\overline F}_a\right]), \quad
[D,L]=0,
\label{suplax}
\end{eqnarray}
where $F_a(Z)$ and ${\overline F}_a(Z)$ ($a,b=1,\ldots , n+m$) are chiral
and antichiral $N=2$ superfields
\begin{eqnarray}
D F_a(Z)=0, \quad {\overline D}~{\overline F}_a(Z) = 0,
\label{chiral}
\end{eqnarray}
respectively, which are bosonic for $a=1,\ldots ,n$ and fermionic for
$a=n+1,\ldots , n+m$; $Z=(z,\theta,\overline\theta)$ is a coordinate of
$N=2$ superspace and $D,{\overline D}$ are the $N=2$ supersymmetric
fermionic covariant derivatives
\begin{eqnarray}\label{DD}
D=\frac{\partial}{\partial\theta}
 -\frac{1}{2}\overline\theta\frac{\partial}{\partial z}, \quad
{\overline D}=\frac{\partial}{\partial\overline\theta}
 -\frac{1}{2}\theta\frac{\partial}{\partial z}, \quad
D^{2}={\overline D}^{2}=0, \quad
\left\{ D,{\overline D} \right\}= -\frac{\partial}{\partial z}
\equiv -{\partial}.
\end{eqnarray}
For positive-integer $k$, such a Lax operator provides the consistent flows
\begin{eqnarray}
{\textstyle{\partial\over\partial t_k}}L =[ A , L], \quad
A=(L^k)_{\geq 1}
\label{laxfl1}
\end{eqnarray}
and the infinite number of conserved currents can be obtained as follows:
\begin{eqnarray}
{H}_k = \int d Z (L^k)_{0},
\label{res}
\end{eqnarray}
where the subscripts $\geq 1$ and $0$ mean the sum of the purely
derivative terms and the constant part of the operator, respectively.
There are four additional integrals of motion
\begin{eqnarray}
\tilde{H}_1 = \int d z F_a {\overline F}_a,
\label{nres}
\end{eqnarray}
where we have only space integration due to the equation of motion
\begin{eqnarray}
-\frac{1}{2}{\textstyle{\partial\over\partial t_k}}
(F_a {\overline F}_a)=((L^k)_0)~',
\label{nnn1}
\end{eqnarray}
where the sign $'$ means the derivative with respect to z.

Equations belonging to the $N=2$ supersymmetric $(n,m)$-GNLS hierarchy
admit the complex structure
\begin{eqnarray}
F_a^{*}= (-i)^{d_a-1} {\cal P}_{ab}{\overline F}_b, \quad
{\overline F_a^{*}}=(-i)^{d_a-1}{\cal P}_{ab} F_b, \quad
{\theta}^{*}={\overline {\theta}}, \quad
{\overline {\theta}^{*}}={\theta}, \quad
t^{*}_k= (-1)^{k+1} t_k, \quad
z^{*}= z,
\label{conj}
\end{eqnarray}
where $i$ is the imaginary unity and $d_a$ define the grading
$F_aF_b=(-1)^{d_ad_b}F_bF_a$ with the property $d_a=1$ $(d_a=0)$ for
fermionic (bosonic) superfields; ${\cal P}_{ab}$ is a permutation matrix
(${\cal P}^2=I$) belonging to the discrete permutation subgroup of the
$GL(n|m)$ supergroup, which is the group of invariance of the Lax operator
\p{suplax}.

 From eq. \p{laxfl1} with the Lax operator \p{suplax}, one
can easily extract the equations for the superfields $F_a$,
\begin{eqnarray}
{\textstyle{\partial\over\partial t_k}}F_a = ((L^k)_{\geq 1} F_a)_0.
\label{n1}
\end{eqnarray}
Applying the transformations \p{conj} to eqs. \p{n1}, one can derive
the corresponding equations for the superfields $\overline F_a$,
\begin{eqnarray}
{\textstyle{\partial\over\partial t_k}}\overline F_a =(-1)^{k+1}
((L^{*~k})_{\geq 1} \overline F_a)_0,
\label{nn1}
\end{eqnarray}
where $L^*$ is the complex-conjugate Lax operator
\begin{eqnarray}
L^*= \partial + \frac{1}{2}(F_a {\overline F}_a -
\overline F_a D\partial^{-1} \left[ {\overline D} F_a\right]), \quad
A^*=(L^{*~k})_{\geq 1}, \quad
[\overline D, L^*]=0,
\label{suplaxconj}
\end{eqnarray}
which also provides the consistent flows.
The first nontrivial flow from \p{n1}, \p{nn1} is the
second flow which reads
\begin{eqnarray}
{\textstyle{\partial\over\partial t_2}} F_a =
F_a~'' +  {D}(F_b {\overline F}_b~{\overline D} F_a), \quad
{\textstyle{\partial\over \partial t_2}} {\overline F}_a =
-{\overline F}_a~'' + {\overline D}
(F_b {\overline F}_b {D}{\overline F}_a).
\label{gnls}
\end{eqnarray}
The set of equations \p{gnls} form the $N=2$ supersymmetric GNLS
equations.

{}~

\noindent{\bf 2. Discrete symmetries of the
$N=2$ super-GNLS hierarchies.}
Here, we demonstrate that in addition to the transformations of the
$N=2$ supersymmetry and $GL(n|m)$ supergroup, the $N=2$ supersymmetric
$(n,m)$-GNLS hierarchy is invariant with respect to discrete mappings.
In the particular cases corresponding to $n=0,m=1$ and $n=1,m=0$, such
mappings were obtained in \cite{dls,s}. Following the scheme
developed in \cite{s}, we derive their generalizations for arbitrary
values of the discrete parameters $n$ and $m$.

Applying the gauge transformation
\begin{eqnarray}
\widetilde{L} = G^{-1} L G, \quad
\widetilde{A} = G^{-1} A G -
G^{-1} {\textstyle{\partial\over\partial t_k}} G,
\quad {\textstyle{\partial\over\partial t_k}}\widetilde{L} =
[\widetilde{A} , \widetilde{L}]
\label{n2}
\end{eqnarray}
with the gauge function $G$ equal to some given bosonic superfield $F_l$,
\begin{eqnarray}
G= F_l
\label{g1}
\end{eqnarray}
(i.e., the index $l$ is an arbitrary fixed index belonging to the range $1
\leq l \leq n$), substituting the $t_k$-derivative of $F_l$ \p{n1} into
\p{n2}, introducing the new superfield basis $\\$
$\{J(Z), {\widetilde{F_j}}(Z), {\widetilde{{\overline F}_j}}(Z), j=1,
\ldots , l-1, l+1, \ldots , n, \ldots , n+m\}$
\begin{eqnarray}
\widetilde{F_j}=\frac{1}{\sqrt{2}}{F_l}^{-1}F_{j}, \quad
\widetilde{{\overline F}_j}=-
\frac{1}{\sqrt{2}}{\overline D} D {\partial}^{-1}(F_l{\overline F}_{j}),
\quad
J=\frac{1}{2}(\frac{1}{2} F_a {{\overline F}_a} - (\ln F_l)~'~)
\label{n3}
\end{eqnarray}
and making obvious algebraic manipulations in the result, we obtain the
following explicit expressions for the operators $\widetilde{L}$ and
$\widetilde{A}$:
\begin{eqnarray}
\widetilde{L}= \partial -
2J-2 {\overline D}\partial^{-1} \left[ D (J -
\frac{1}{2}{\widetilde F_j} {\widetilde {\overline F}_j})
\right]- {\widetilde F_j} {\overline D}\partial^{-1}
\left[ D {\widetilde {\overline F}_j}\right], \quad
\widetilde{A}=(\widetilde{L}^k)_{\geq 1},
\label{n4}
\end{eqnarray}
which coincide with the LA-pair considered in \cite{ik}.
Thus, the integrable extension of the $N=2$ supersymmetric $a=4$ KdV
hierarchy of Ref. \cite{ik} is gauge related to the $N=2$
supersymmetric $(n,m)$-GNLS hierarchy \cite{bks} and relations \p{n3}
establish their explicit connection. For the particular case $n=1, m=0$,
relation \p{n3} was obtained in \cite{bks}.

In the new basis \p{n3}, the second flow equations \p{gnls} become
\begin{eqnarray}
&&{\textstyle{\partial\over\partial t_2}}{\widetilde F_j} =
{\widetilde F_j}~''+ 4D(J\overline D {\widetilde F_j}), \quad
{\textstyle{\partial\over\partial t_2}}{\widetilde {\overline F}_j} =
-{\widetilde {\overline  F}_j}~''+
4\overline D (J D {\widetilde {\overline F}_j}), \nonumber\\
&&{\textstyle{\partial\over\partial t_2}}J = (-[D,{\overline D}] J-2J^2+
D{\widetilde {\overline F}_j} \cdot \overline D {\widetilde F_j})~',
\label{ngnls}
\end{eqnarray}
and one can observe that they, as well as other equations belonging to the
hierarchy, admit the complex structure
\begin{eqnarray}
&&{\widetilde F_j^{*}}= (-i)^{d_j-1}\widetilde {\cal P}_{jc}
{\widetilde {\overline F}_c}, \quad
{\widetilde {\overline F}_j^{*}}=(-i)^{d_j-1}\widetilde {\cal P}_{jc}
{\widetilde F_c}, \quad
J^{*} = -J, \nonumber\\
&&{\theta}^{*}={\overline {\theta}}, \quad
{\overline {\theta}^{*}}={\theta}, \quad
t^{*}_k= (-1)^{k+1} t_k, \quad
z^{*}= z.
\label{nconj}
\end{eqnarray}
Applying the complex-conjugation transformations \p{conj} and \p{nconj} to
\p{n3}, we observe that in addition to relation \p{n3}, there exists one
more
\begin{eqnarray}
\widetilde{F_j}= -\frac{1}{\sqrt{2}}i D {\overline D} {\partial}^{-1}
({\overline F}_l{\cal P}_{jc} F_{c}), \quad
\widetilde{{\overline F}_j}=-\frac{1}{\sqrt{2}}i{{\overline F}_l}^{-1}
{\cal P}_{jc} {\overline F}_{c}, \quad
J=\frac{1}{2}(\frac{1}{2} F_a {{\overline F}_a} + (\ln {\overline F}_l)~'~),
\label{n5}
\end{eqnarray}
which connects the second flow equations \p{ngnls} to \p{gnls}, as well as
their corresponding hierarchies. Denoting the superfields $F_a$ and
$\overline F_a$ in \p{n3} by the new letters $\r F{}_a$ and $\r {\overline
F}{}_a$, respectively, and equating the corresponding superfields
${\widetilde F_j}$, ${\widetilde {\overline F}_j}$ and $J$ belonging to
the relations \p{n3} and \p{n5}, we derive the mapping
\begin{eqnarray}
&& D {\overline D}{\partial}^{-1}({\overline F}_l F_{j})=
i{\r F{}_l}^{-1}{\cal P}_{jc} \r F{}_{c}, \quad
{\overline D} D {\partial}^{-1} (\r F{}_l\r {\overline F}{}_{j})=
i{{\overline F}_l}^{-1}{\cal P}_{jc} {\overline F}_{c}, \nonumber\\
&&\frac{1}{2} (\r F{}_l \r {\overline F}{}_l +
\r F{}_{j} \r {\overline F}{}_{j} - F_l {{\overline F}_l} -
F_{j} {{\overline F}_{j}}) = (\ln (\r F{}_l {\overline F}_l))~'
\label{n6}
\end{eqnarray}
that acts like the discrete symmetry transformation of the $N=2$
supersymmetric $(n,m)$-GNLS hierarchy. Acting by the fermionic covariant
derivatives $D$ and $\overline D$ on the first line of eqs. \p{n6},
these relations can be rewritten in a slightly different but equivalent
local form
\begin{eqnarray}
{\overline D}~({\overline F}_l F_{j}+
i{\r F{}_l}^{-1}{\cal P}_{jc} \r F{}_{c})=0, \quad
D~ (\r F{}_l\r {\overline F}{}_{j}+
i{{\overline F}_l}^{-1}{\cal P}_{jc} {\overline F}_{c})=0,
\label{nnn6}
\end{eqnarray}
which can be more convenient for applications. Actually, it is easy to
understand that up to an arbitrary permutation ${\cal P}_{jc}$, relation
\p{n6} gives us $n$ different discrete symmetry-mappings if one
remembers that the index $l$ enters \p{n6} like a discrete
parameter\footnote{Let us remember that in \p{n6}, there is no summation
over repeated indices $l$.} taking $n$ values $l=1, \ldots , n$.

Now consider gauge transformation \p{n2} with the gauge function
\begin{eqnarray}
G = (D {\overline F}_f)^{-1},
\label{g2}
\end{eqnarray}
where $\overline F_f$ is some given fermionic superfield
(i.e., the index $f$ is an arbitrary fixed index belonging to the range
$n+1 \leq f \leq n+m$). After introducing the
new superfield basis  $\{J(Z), {\widetilde{F_j}}(Z),
{\widetilde{{\overline F}_j}}(Z),j=1, \ldots , n, \ldots ,
f-1, f+1, \ldots , n+m\}$
according to formulae
\begin{eqnarray}
\widetilde{F_j}=
\frac{1}{\sqrt{2}}(D {\overline F}_f)F_{j}, \quad
\widetilde{{\overline F}_j}=-
\frac{1}{\sqrt{2}}{\overline D}
D ((D {\overline F}_f)^{-1}{\overline F}_{j}),\quad
J=\frac{1}{2}(\frac{1}{2} F_a {{\overline F}_a} +
(\ln D {\overline F}_f)~'~)
\label{nn3}
\end{eqnarray}
we obtain the following expression for the Lax operator
$\widetilde{L}$
\begin{eqnarray}
\widetilde{L}= \partial - 2J+2\left[ D (J - \frac{1}{2}{\widetilde F_j}
\partial^{-1} {\widetilde {\overline F}_j})\right]
{\overline D}\partial^{-1} - {\widetilde F_j} {\overline D}\partial^{-1}
\left[ D \partial^{-1} {\widetilde {\overline F}_j}\right].
\label{nn4}
\end{eqnarray}
We do not present the explicit expression for the operator
$\widetilde{A}$ here, because what we actually need for our purpose is
only the transformation law \p{nn3} in the new basis. For the particular
case $n=0, m=1$, relation \p{nn3} has been discussed in \cite{ks,kst}.

In the new basis \p{nn3}, the second flow equations \p{gnls} become
\begin{eqnarray}
&&{\textstyle{\partial\over\partial t_2}}{\widetilde F_j} =
{\widetilde F_j}~''+ 4D\overline D (J{\widetilde F_j}), \quad
{\textstyle{\partial\over\partial t_2}}{\widetilde {\overline F}_j} =
-{\widetilde {\overline  F}_j}~''+
4\overline D D(J {\widetilde {\overline F}_j}), \nonumber\\
&&{\textstyle{\partial\over\partial t_2}}J =( [D,{\overline D}] J-2J^2-
{\widetilde F_j} {\widetilde {\overline F}_j})~',
\label{nngnls}
\end{eqnarray}
and one can see that they admit the complex structure
\begin{eqnarray}
&&{\widetilde F_j^{*}}= (-i)^{d_j} {\widetilde {\cal P}}_{jc}{\widetilde
{\overline F}_c}, \quad
{\widetilde {\overline F}_j^{*}}=(-i)^{d_j}{\widetilde {\cal P}}_{jc}
{\widetilde F_c}, \quad J^{*} = -J, \nonumber\\
&& {\theta}^{*}={\overline {\theta}}, \quad
{\overline {\theta}^{*}}={\theta}, \quad
t^{*}_2= -t_2, \quad
z^{*}= z.
\label{nnconj}
\end{eqnarray}

Following the above-discussed scheme, we apply the
complex-conjugation transformations \p{conj} and \p{nnconj} to \p{nn3} and
obtain one more mapping,
\begin{eqnarray}
\widetilde{F_j}=
-\frac{i}{\sqrt{2}} D {\overline D}
(({\overline D} F_f)^{-1}{\cal P}_{jc} F_{c}), \quad
\widetilde{{\overline F}_j}=
\frac{i}{\sqrt{2}}({\overline D}F_f){\cal P}_{jc} {\overline F}_{c}, \quad
J=\frac{1}{2}(\frac{1}{2} F_a {{\overline F}_a} -
(\ln {\overline D} F_f)~'~),
\label{nn5}
\end{eqnarray}
connecting the second flow equations \p{nngnls} to \p{gnls}, and,
therefore, the mapping
\begin{eqnarray}
&& i D {\overline D}((\overline D F_f)^{-1} F_{j})=
-(D \r {{\overline F}}{}_f){\cal P}_{jc} \r F{}_{c},\quad
i{\overline D} D ((D \r {{\overline F}}{}_f)^{-1}
\r {\overline F}{}_{j})=
(\overline D F_f){\cal P}_{jc} {\overline F}_{c}, \nonumber\\
&&\frac{1}{2} (F_f {{\overline F}_f} + F_{j} {{\overline F}_{j}}-
\r F{}_f \r {\overline F}{}_f -\r F{}_{j} \r {\overline F}{}_{j}) =
(\ln (\overline D F_f \cdot D \r {{\overline F}}{}_f))~'
\label{nn6}
\end{eqnarray}
acts like the discrete symmetry transformation of the $N=2$ supersymmetric
$(n,m)$-GNLS hierarchy. Acting by the fermionic covariant derivatives $D$
and $\overline D$ on the first line of eqs. \p{nn6},
these relations can be represented in the following equivalent form:
\begin{eqnarray}
{\overline D}~ (-i((\overline D F_f)^{-1} F_{j})~' +
(D \r {{\overline F}}{}_f){\cal P}_{jc} \r F{}_{c})=0, \quad
D~ (i((D \r {{\overline F}}{}_f)^{-1} \r {\overline F}{}_{j})~'+
(\overline D F_f){\cal P}_{jc} {\overline F}_{c})=0.
\label{nnnn6}
\end{eqnarray}
Modulo an arbitrary permutation ${\cal P}_{jc}$, relation \p{nn6} gives us
$m$ different discrete symmetry-mappings because the index $f$ takes $m$
different values $l=n+1, \ldots , n+m$.

Let us note that one can rewrite equations (\ref{nngnls}) in a form
similar to (\ref{ngnls}),
\begin{eqnarray}
&&-{\textstyle{\partial\over\partial t_2}}{\Psi_j} =
{\Psi_j}~''+ 4D({\widetilde J}~\overline D {\Psi_j}), \quad
-{\textstyle{\partial\over\partial t_2}}{\overline \Psi}_j =
-{\overline  \Psi}_j~''+
4\overline D ({\widetilde J} D {\overline \Psi}_j), \nonumber\\
&&-{\textstyle{\partial\over\partial t_2}}{\widetilde J} =
(-[D,{\overline D}] {\widetilde J}-2{\widetilde J}^2+
D{\overline \Psi}_j \cdot \overline D {\Psi_j})~',
\label{nnngnls}
\end{eqnarray}
if one introduces the new superfields ${\widetilde J}$, $\Psi_j$ and
${\overline \Psi}_j$ by the following invertible
relations\footnote{Transformations of such kind have been discussed in
\cite{ik}.}
\begin{eqnarray}
&& {\widetilde J}=-J, \quad
\Psi_j = i {\partial}^{-1} D {\widetilde {\overline F}_j}, \quad
{\overline \Psi_j} =i{\partial}^{-1} {\overline D} {\widetilde F_j};
\nonumber\\
&& J=-{\widetilde J}, \quad
{\widetilde F_j} = i D {\overline \Psi}_j, \quad
{\widetilde {\overline F}_j} = i {\overline D} \Psi_j.
\label{tr}
\end{eqnarray}
However the system (\ref{nnngnls}) does not completely coincide with
(\ref{ngnls}): in comparison with (\ref{ngnls}), its time direction is
reversed. Due to this crucial difference, we can not equate the
corresponding superfields entering (\ref{ngnls}) and (\ref{nnngnls}) to
produce new discrete symmetry-mappings for the $N=2$ supersymmetric
$(n,m)$-GNLS hierarchy. Nevertheless, the system (\ref{nnngnls}) is
equivalent to (\ref{ngnls}), and relations \p{nn3}, \p{nn5}, and
\p{tr} establish an explicit connection of the integrable hierarchy of
Ref. \cite{ik} to the $N=2$ supersymmetric $(n,m)$-GNLS hierarchy.

{}~

\noindent{\bf 3. Bosonic limit of the mappings.}
Let us briefly discuss the bosonic limit of the mappings \p{n6} and
\p{nn6} in order to generate the discrete symmetries for the
bosonic GNLS and modified GNLS (mGNLS) hierarchies.

To do this, we set all fermionic components of the superfields $F_a$
and $\overline F_a$ equal to zero and define the bosonic
components as \cite{bks}
\begin{eqnarray}
&& b_{\alpha}  = \frac{1}{\sqrt{2}} F_{\alpha}|, \quad
{\overline b}_{\beta} = \frac{1}{\sqrt{2}} {\overline F}_{\beta}|, \quad
1 \leq {\alpha},{\beta} \leq n, \nonumber\\
&& g_s = \frac{1}{\sqrt{2}} {\overline D} F_{s+n}|~
{\exp (-{\partial^{-1}} (b_{\beta} {\overline b}_{\beta}))}, \quad
{\overline g}_p = \frac{1}{\sqrt{2}} D {\overline F}_{p+n}|~
{\exp ({\partial^{-1}} (b_{\beta} {\overline b}_{\beta}))}, \quad
1 \leq s,p \leq m, ~ ~ ~ ~ ~
\label{def}
\end{eqnarray}
where $|$ means the $({\theta}, {\bar\theta})\rightarrow 0$ limit.
In terms of such components, eqs. \p{gnls} for the fields
$b_{\alpha},{\overline b}_{\alpha}$ and $g_s, \overline g_s$ are
completely decoupled:
\begin{eqnarray}
{\textstyle{\partial\over\partial t_2}} b_{\alpha}=
b_{\alpha}~'' -  2 b_{\beta} {\overline b}_{\beta} b_{\alpha}~', \quad
{\textstyle{\partial\over \partial t_2}} {\overline b}_{\alpha} =
-{\overline b}_{\alpha}~'' - 2 b_{\beta}
   {\overline b}_{\beta} {\overline b}_{\alpha}~',
\label{bgnls}
\end{eqnarray}
\begin{eqnarray}
{\textstyle{\partial\over\partial t_2}} g_s=
g_s~'' -  2 g_p {\overline g}_p g_s, \quad
{\textstyle{\partial\over \partial t_2}} {\overline g}_s =
-{\overline g}_s~'' + 2 g_p {\overline g}_p {\overline g}_s.
\label{fgnls}
\end{eqnarray}
The set of equations \p{fgnls} form the bosonic GNLS equations \cite{fk}.
Concerning the set of equations \p{bgnls}, we call them mGNLS
equations, reflecting the name of its first representative---modified NLS
equation \cite{cll} corresponding to the case of $n=1$.

The bosonic limit of the mapping \p{n6} ( \p{nn6} ), acting like
a discrete symmetry transformation of equations \p{bgnls} and \p{fgnls},
also splits into two independent mappings, which one can see from the
explicit expressions
\begin{eqnarray}
&& {\overline b}_l b_{\alpha}~'+
i({\r b{}_l}^{-1}{\cal P}_{{\alpha}{\beta}} \r b{}_{\beta})~' = 0, \quad
\r b{}_l \r {\overline b}{}_{\alpha}~'+
i({{\overline b}_l}^{-1} {\cal P}_{{\alpha}{\beta}}
{\overline b}_{\beta})~'=0, \quad {\alpha} \neq l, \nonumber\\
&& \r b{}_l \r {\overline b}{}_l+
\r b{}_{\alpha} \r {\overline b}{}_{\alpha} -
b_l {\overline b}_l
-b_{\alpha} {{\overline b}_{\alpha}} =
(\ln (\r b{}_l {\overline b}_l))~', \quad
\r b{}_l \r {\overline b}{}_l~'+
\r b{}_{\alpha} \r {\overline b}{}_{\alpha}~' -
b_l {\overline b}_l~'
-b_{\alpha} {\overline b}_{\alpha}~' =
(\ln {\overline b}_l)~'', ~ ~ ~ ~ ~ ~
\label{bn6}
\end{eqnarray}
\begin{eqnarray}
\r g{}_s = {\cal P}_{sp} g_p, \quad
\r {\overline g}{}_s = {\cal P}_{sp} {\overline g}_p
\label{bbn6}
\end{eqnarray}
for the mapping \p{n6}, and
\begin{eqnarray}
\r b{}_{\alpha}{\cal P}_{{\alpha}{\beta}} \r {\overline b}{}_{\beta}~'+
b_{\alpha} {\cal P}_{{\alpha}{\beta}} {\overline b}_{\beta}~'=0, \quad
\r b{}_{\beta} \r {\overline b}{}_{\beta} -
b_{\beta} {{\overline b}_{\beta}} =
-(\ln {\overline b}_{\alpha}~')~' +
(\ln ({\cal P}_{{\alpha}{\beta}} \r {\overline b}{}_{\beta}))~',
\label{fn6}
\end{eqnarray}
\begin{eqnarray}
&& -i(g_f^{-1} g_{s})~' +
\r {{\overline g}}{}_f{\cal P}_{sp} \r g{}_{p}=0, \quad
i(\r {{\overline g}}{}_f^{-1} \r {\overline g}{}_{s})~'+
 g_f {\cal P}_{sp} {\overline g}_{p}=0, \nonumber\\
&& \r {\overline g}{}_f = C_f g_f^{-1}, \quad
\r g{}_f \r {\overline g}{}_f+
\r g{}_s \r {\overline g}{}_s - g_f {\overline g}_f
-g_s {\overline g}_s =
-(\ln {\overline g}_f)~'', \quad s \neq f
\label{ffn6}
\end{eqnarray}
for the mapping \p{nn6}, where there is no summation in eq. \p{fn6} over
repeated indices ${\alpha}$ and $C_f$ in eq. \p{ffn6} is an arbitrary
constant. In the derivation of these expressions, obvious simplifying
transformations, as well as the integration of some intermediate
equations, have been done.

The mapping \p{bbn6} forms the discrete permutation subgroup of the
$GL(m)$ group, which is a group of covariance for the GNLS equations
\p{fgnls}. The mapping \p{ffn6} coincides with the mapping which can be
easily derived using the Darboux-B\"acklund transformations of the GNLS
Lax operators constructed in \cite{a}. Regarding the symmetry
mappings \p{bn6} and \p{fn6} for the mGNLS hierarchy \p{bgnls},
to our knowledge, they are presented for the first time.

In addition to mappings \p{bn6} and \p{fn6}, there are other
symmetries of the mGNLS equations. One can produce them if one remembers
that the GNLS and mGNLS equations are related by the following
transformations \cite{bks}:
\begin{eqnarray}
g_s= b_s~'~ {\exp (-{\partial^{-1}} (b_p {\overline b}_p))}, \quad
{\overline g}_s= {\overline b}_s~ {\exp ({\partial^{-1}}
(b_p {\overline b}_p))}.
\label{trmb}
\end{eqnarray}
Applying the complex-conjugation operation \p{conj} to relations \p{trmb}
for the bosonic components \p{def}, one can obtain one more relation,
\begin{eqnarray}
g_s= i b_s~ {\exp (-{\partial^{-1}} (b_p {\overline b}_p))}, \quad
{\overline g}_s=
i {\overline b}_s~'~{\exp ({\partial^{-1}} (b_p {\overline b}_p))}.
\label{ctrmb}
\end{eqnarray}
Therefore, we can introduce two different relations for the fields
with the arrow:
\begin{eqnarray}
&& \r g{}_s= \r b{}_s~'~ {\exp (-{\partial^{-1}}
(\r b{}_p \r {\overline b}{}_p))}, \quad
\r {\overline g}{}_s=
\r {\overline b}{}_s~ {\exp ({\partial^{-1}}
(\r b{}_p \r {\overline b}{}_p))}, \nonumber\\
&& \r g{}_s= i \r b{}_s~ {\exp (-{\partial^{-1}} (\r b{}_p
\r {\overline b}{}_p))}, \quad
\r {\overline g}{}_s= i \r {\overline b}{}_s~'~{\exp
({\partial^{-1}} (\r b{}_p \r {\overline b}{}_p))}.
\label{atrmb}
\end{eqnarray}
If one takes some fixed combination of the fields without the arrow, $g_s,
\overline g_s$, and the fields with the arrow $\r g{}_s, \r {\overline
g}{}_s$ from the set of relations \p{trmb}--\p{atrmb}, and substitutes
it into the mappings \p{bbn6} and \p{ffn6}, one can generate new
mappings for the mGNLS hierarchies, with different
combinations generating different mappings. Let us only mention that one
such mapping coincides with the mapping \p{fn6}, and, in this way, it is
possible to reproduce the mapping considered in \cite{l} for the modified
NLS equation and to obtain new mappings. It is a simple exercise to derive
their explicit forms, and we do not present them here.

{}~

\noindent{\bf 4. Conclusion.}
In this Letter, we constructed mappings \p{n6} and \p{nn6}, which act
like a discrete symmetry transformations of the $N=2$ supersymmetric
$(n,m)$-GNLS hierarchy \p{suplax}, \p{laxfl1}, and produced their bosonic
counterparts \p{bn6}--\p{ffn6}. We also established explicit relations
\p{n3}, \p{n5}, \p{nn3}, and \p{nn5} connecting the integrable hierarchy,
obtained by the junction of the Lax operators for the $N=2$ supersymmetric
$a=4$ KdV and $(n-1,m)$-GNLS hierarchies, to the $N=2$ supersymmetric
$(n,m)$-GNLS hierarchy.

Symmetry mappings contain valuable information about the integrable
hierarchies corresponding to them \cite{lf,ls1,dls,s}. In addition to this,
there is one more reason that stimulates interest in such
mappings---they are usually integrable themselves, i.e., every new mapping
may give us a new example of a one-dimensional integrable system. Thus,
for the $N=2$ supersymmetric f-Toda chain \cite{dls} corresponding to the
case of $n=0, m=1$, the integrability under appropriate boundary
conditions has been proven in \cite{ls}. It is interesting to generalize
this investigation for the case of arbitrary values of the discrete
parameters $n$ and $m$. In this context, it is important to have
Darboux-B\"acklund transformations of the $N=2$ supersymmetric
$(n,m)$-GNLS Lax operators which should generate our mappings and contain
important information about their integrability properties and solutions.
We hope to analyze this complicated problem in future publications.

{}~

\noindent{\bf Acknowledgments.} I would like to thank L. Bonora,
E.A. Ivanov, A.N. Leznov and G.M. Zinovjev for their interest in this work
and useful discussions. This work was partially supported by the Russian
Foundation for Basic Research, Grant No. 96-02-17634, INTAS Grant No.
94-2317, and by a grant from the Dutch NWO organization.

\end{document}